# Demonstration of High $f_{max}$ (> 80 GHz) in Recessed-Gate and Passivated $Al_{0.6}Ga_{0.4}N$ Channel RF HEMTs

Meherab Hossain*, Ruixin Bai, Swarnav Mukhopadhyay, Brahmani Challa, Guangying Wang, Surjava Sanyal, Md Tahmidul Alam, Khush Gohel, Rafael A. Choudhury, Shubhra S. Pasayat, and Chirag Gupta

*Department of Electrical and Computer Engineering, University of Wisconsin-Madison, WI 53706, USA*

*E-mail: meherab.hossain@wisc.edu

In this work, ultra-wide bandgap (UWBG) $Al_{0.60}Ga_{0.40}N$ channel high-electron-mobility transistors (HEMTs) with an $Al_{0.81}Ga_{0.19}N$ barrier and an $Al_{0.86}Ga_{0.14}N$ back barrier were fabricated. The devices featured a recessed T-gate structure and $SiO_2$ passivation. Scaled devices achieved a cutoff frequency ($f_T$) of 35 GHz, a maximum oscillation frequency ($f_{max}$) of 82 GHz, and an average breakdown field > 3 MV/cm, while maintaining a reasonable current collapse of ~20% under a 40 V drain bias in pinch-off condition. Devices with larger dimensions demonstrated a high off-state breakdown voltage of 883 V and an $f_T$ of 24 GHz, yielding a high Johnson's figure of merit (JFOM) of 21.19 THz·V. Across all device geometries, a high $I_{ON}/I_{OFF}$ ratio exceeding $10^6$ was observed with drain and gate leakage currents strictly below $10^{-6}$ A/mm. This work demonstrates an optimized design for AlGaN channel RF HEMTs and their excellent potential for X- to Ka-band applications.

GaN HEMTs have demonstrated excellent high power and high frequency performance. While GaN HEMTs continue to improve, the critical electric field may eventually limit the overall device performance. In this regard, with a high critical electric field [1], [2], [3], wider bandgap, and comparable saturation velocity (compared to GaN) [4], [5], AlGaN channel transistors have the potential to provide much higher power for high-frequency applications as well as extreme environment operation capability [6], [7].

Consequently, $Al_xGa_{1-x}N$ (x>0.5) channel FETs have gained attention and have been explored for both power [8], [9], [10], [11], [12], [13] and RF applications [14], [15], [16], [17], [18]. So far, the highest reported $f_{max}$ in AlGaN channel FETs is ~78 GHz with an $f_T$ of ~24 GHz [15]. Also, a high $f_T$ of 85 GHz has been demonstrated, however, the $f_{max}$ was limited to ~20 GHz [16]. In both these reports, passivation was not present. In a practical device, passivation can be expected to manage current collapse or dispersion. Therefore, significant further improvements in device performance are needed to enable high-frequency performance in AlGaN channel FETs.

To obtain high-frequency operation, simultaneous lateral and vertical scaling is needed. While lateral scaling (i.e. gate length reduction) provides higher cutoff frequency ($f_T$) by reducing the transit time under the gate, it may also deteriorate the aspect ratio without appropriate vertical scaling. Vertical scaling in AlGaN channel FETs is more challenging than typical GaN HEMTs.

The barrier thickness (or gate to channel distance) in AlGaN channel FETs is ~20-25nm [15], [16], [17], [18], [19], [20]. There are two main reasons for this thicker barrier and difficulty in scaling this to < 10-15 nm: 1) The charge density and sheet resistance under the gate as well as in the access regions is directly proportional to the barrier thickness [19]. Therefore, vertical scaling can increase access region sheet resistance. 2) The reverse grade layer etch tolerance [21]. In AlGaN channel FETs, reverse grade layer needs to be etched in the source drain access region, however, due to time control nature of the etch, enough margin is needed to avoid high resistance in the source-drain access region. Therefore, it is non-trivial to perform simultaneous lateral and vertical scaling in AlGaN channel HFETs.

To overcome this fundamental bottleneck, we propose recessed-gate AlGaN channel HEMTs. This design helps in decoupling the access region sheet resistance and aspect ratio. This design has the potential to enable relatively high $f_T$ and $f_{max}$ simultaneously. With the proposed recessed-gate design, even with passivation, ~50% improvement in peak $f_T$ (from 24 GHz to 35 GHz) was obtained compared to the previous design with the highest reported $f_{max}$ for $Al_xGa_{1-x}N$ (x>0.5) channel FETs (~78 GHz) [15]. A simultaneous improvement in $f_{max}$ was also obtained from ~78 GHz to ~82 GHz. High average breakdown fields >3 MV/cm were also obtained for these HEMTs.

While maintaining a high $f_T$ and $f_{max}$ is critical for RF applications, high breakdown voltage is also important for applications at higher power. Johnson's figure of merit (JFOM) has been widely used to evaluate device performance for high power and high frequency applications as this shows the trade-off between achievable frequency and the power handling capability of a device. To date, the highest JFOM reported for a passivated RF HEMT is 20.92 THz·V, achieved in an AlGaN channel device on bulk AlN substrate [17]. In this work, with a recessed-gate structure and optimized device dimensions, an $f_T$ of 24 GHz was achieved with a high off-state breakdown voltage of 883 V, resulting in a high JFOM of 21.19 THz·V on sapphire substrate. Additionally, reasonable dispersion control was achieved across all device dimensions, with both off-state drain and gate leakage currents strictly suppressed to the measurement noise floor.

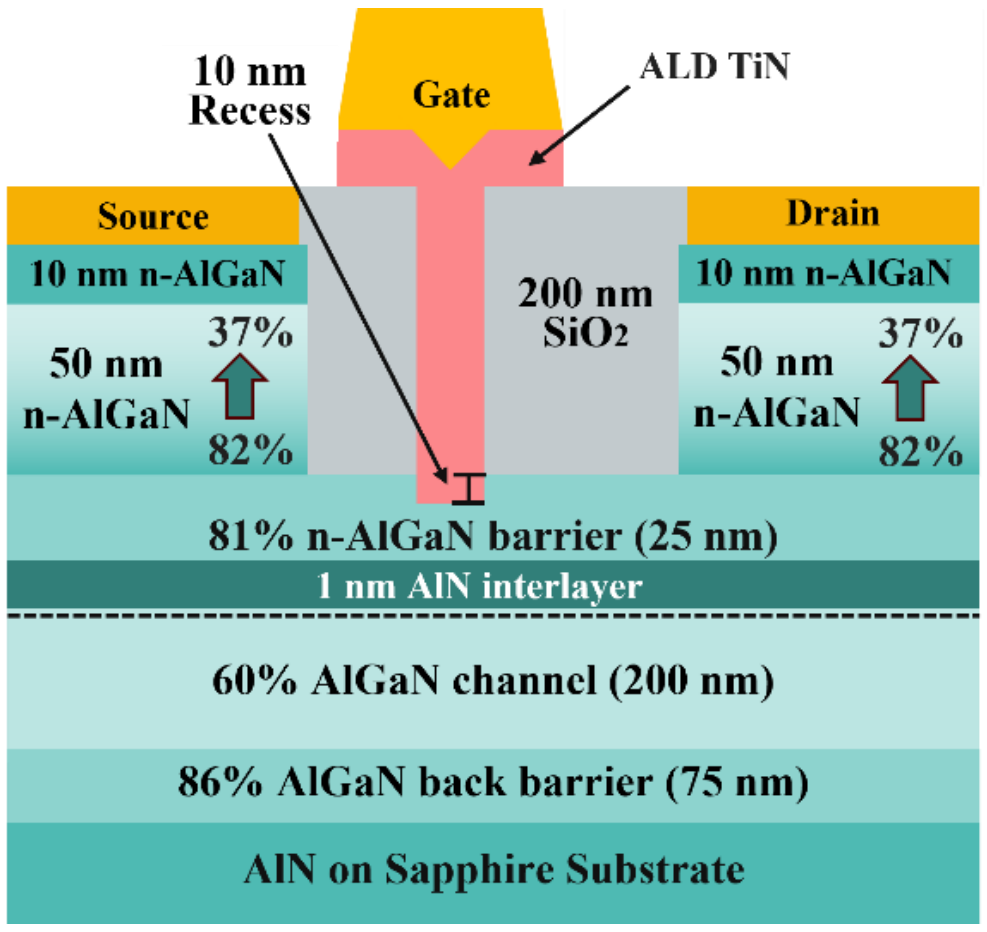

**Figure 1.** Schematic cross section of the passivated gate-recessed $Al_{0.6}Ga_{0.4}N$ channel RF HEMT.

The epitaxial structure used in this work, shown in Fig. 1, was grown in an MOCVD reactor on an AlN on sapphire substrate. The epitaxial structure contained a 10 nm n-AlGaN contact layer, a 50 nm reverse graded layer (Al composition graded from 82% to 37%), a 25 nm $Al_{0.81}Ga_{0.19}N$ barrier, a 1 nm AlN interlayer, a 200 nm $Al_{0.60}Ga_{0.40}N$ channel, and a 75 nm $Al_{0.86}Ga_{0.14}N$ back barrier. More details on the epitaxial growth can be found in [22].

Device fabrication started with a Ti/Al/Ni/Au (20/120/30/50 nm) deposition, followed by annealing in nitrogen ($N_2$) at 1000°C for 30 s. The reverse graded layer was removed using low-power chlorine ($Cl_2$) based time-controlled reactive ion etch (RIE), followed by a deep mesa isolation of 115 nm. A 200 nm thick $SiO_2$ layer was deposited using plasma-enhanced chemical vapor deposition (PECVD) at 350°C. The gate was patterned with electron beam lithography (EBL), and the $SiO_2$ layer was etched in $CHF_3/CF_4/O_2$ plasma. Later, a ~10 nm gate recess was performed using time-controlled low-power $Cl_2$-based RIE. TiN was deposited using ALD (500 cycles) at 275°C. Then, the top-gate was patterned using EBL, followed by a Cr/Au (35/365 nm) metal stack. Later, the via openings were created with the removal of $SiO_2$ using a $CF_4/O_2$ etch. Finally, a Ti/Au (40/460 nm) metal stack was deposited to create probe pads.

To systematically evaluate the recessed-gate structure for high-frequency and high-power performance, a series of devices were fabricated with varying dimensions. The source-to-drain spacing ($L_{sd}$) was varied from 1.0 µm to 6.0 µm, the gate length ($L_g$) from 110 nm to 330 nm, and gate-to-source distance ($L_{sg}$) from 0.35 µm to 0.60 µm. All devices had a total gate width ($W_g$) of 75 µm (2×37.5 µm). Unless otherwise noted, the static DC I-V, pulsed I-V, transfer I-V, and small-signal characteristics presented initially in the following section correspond to the device exhibiting peak high-frequency performance (device A). This specific device had a tightly scaled access region with an $L_{sd}$ of 1.0 µm, an $L_{sg}$ of 0.35 µm, and an $L_g$ of 110 nm. The highest JFOM was achieved by a device with a larger access region (device B), which contained an $L_{sd}$ of 6.0 µm, an $L_{sg}$ of 0.60 µm, and an $L_g$ of 110 nm. Discussion on this device is presented later in the following section.

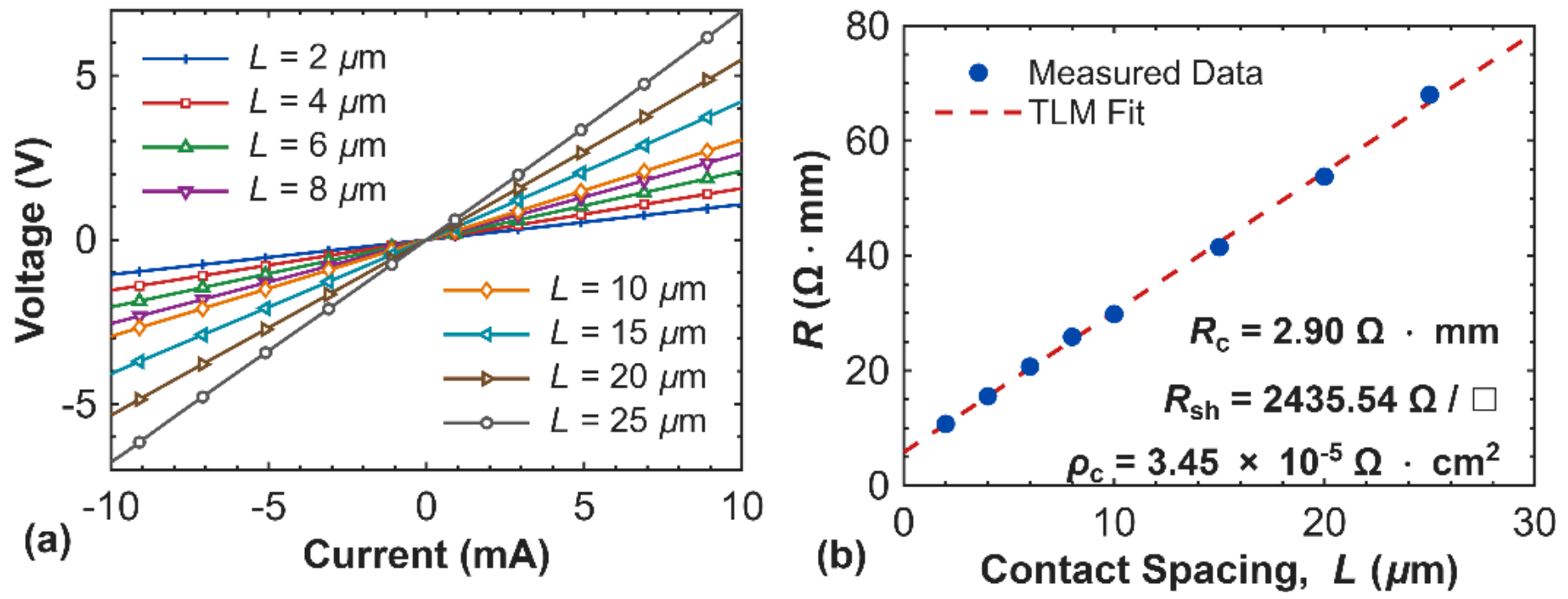

**Figure 2.** (a) I-V curves measured for different TLM spacing (b) Measured resistance and Linear fit of the TLM data.

Fig. 2(a) shows the data from transmission line measurements (TLMs), which demonstrated linear contact. A contact resistance ($R_c$) of 2.90 Ω·mm and a sheet resistance ($R_{sh}$) of 2435.54 Ω/□ were extracted from the TLM data shown in Fig. 2(b).

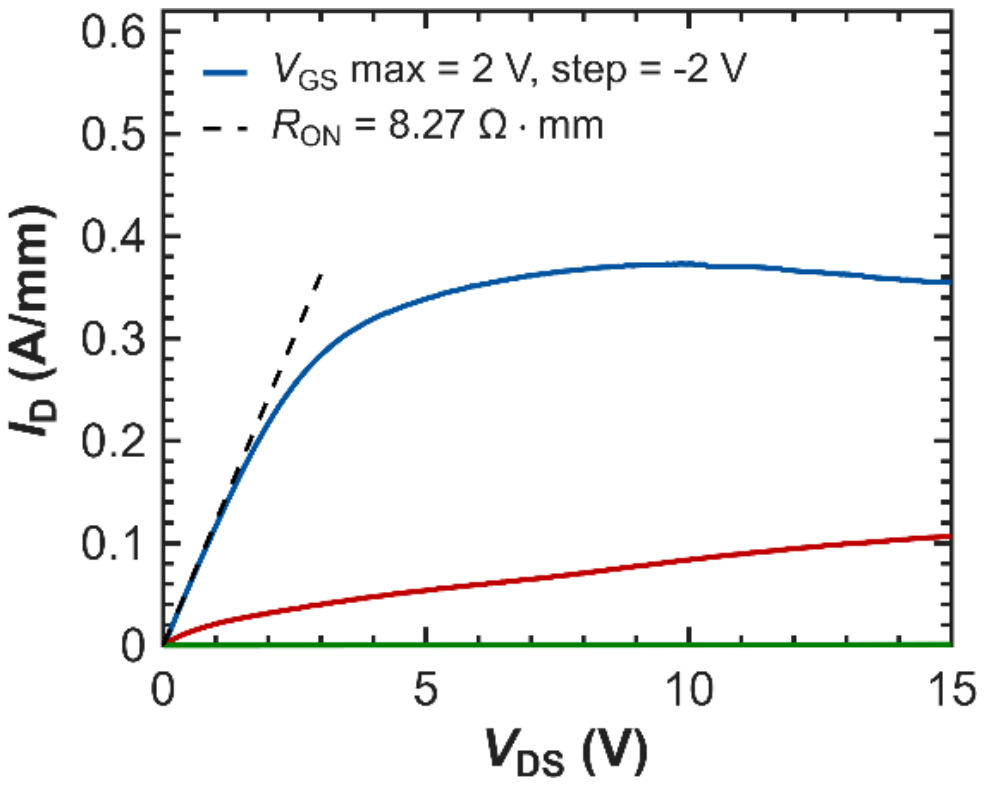

**Figure 3.** Static I-V characteristics of device A.

From the static output IV characteristics of device A shown in Fig. 3, a peak drain current density ($I_D$) of 373 mA/mm was achieved at $V_{GS}$ = 2V. The device demonstrated a low on-resistance ($R_{on}$) of 8.27 Ω·mm, which closely matched the TLM data.

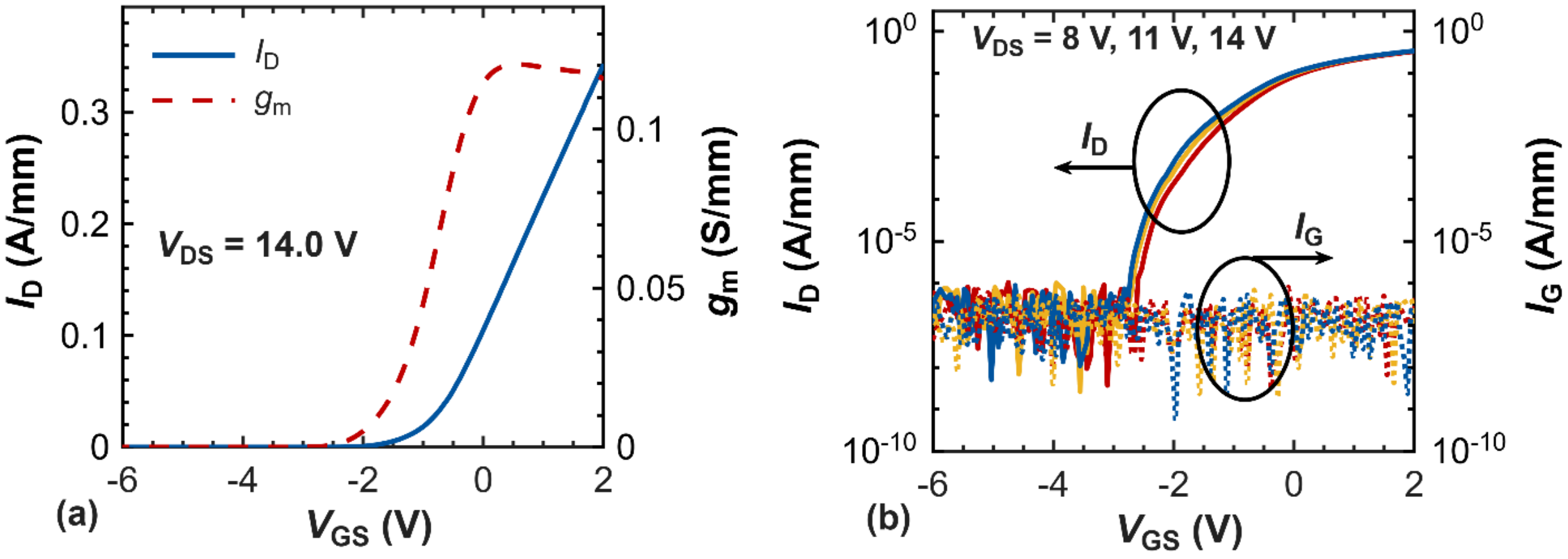

**Figure 4.** Transfer characteristics of device A in (a) linear scale at $V_{DS}$ = 14 V, and (b) log scale at $V_{DS}$ = 8 V (red), 11 V (yellow) and 14 V (blue).

The transfer characteristics of the device on linear scales are shown in Fig. 4(a). A high peak transconductance ($g_m$) of 120 mS/mm was achieved at $V_{GS}$ = 0.53 V for $V_{DS}$ = 14 V. This is one of the highest DC transconductance values achieved for AlGaN channel FETs so far [16], [17], [18], [19], [20], [23], [24]. Fig. 4(b) shows drain current ($I_D$) and gate current ($I_G$) on log scales. The device demonstrated a threshold voltage ($V_{th}$) of -2 V for $V_{DS}$ = 14 V, defined at $I_D = 10^{-3}$ A/mm. The device showed a low gate and drain leakage of ~$10^{-7}$ A/mm. It should be noted that both the drain and gate currents reached the noise floor of the instrument.

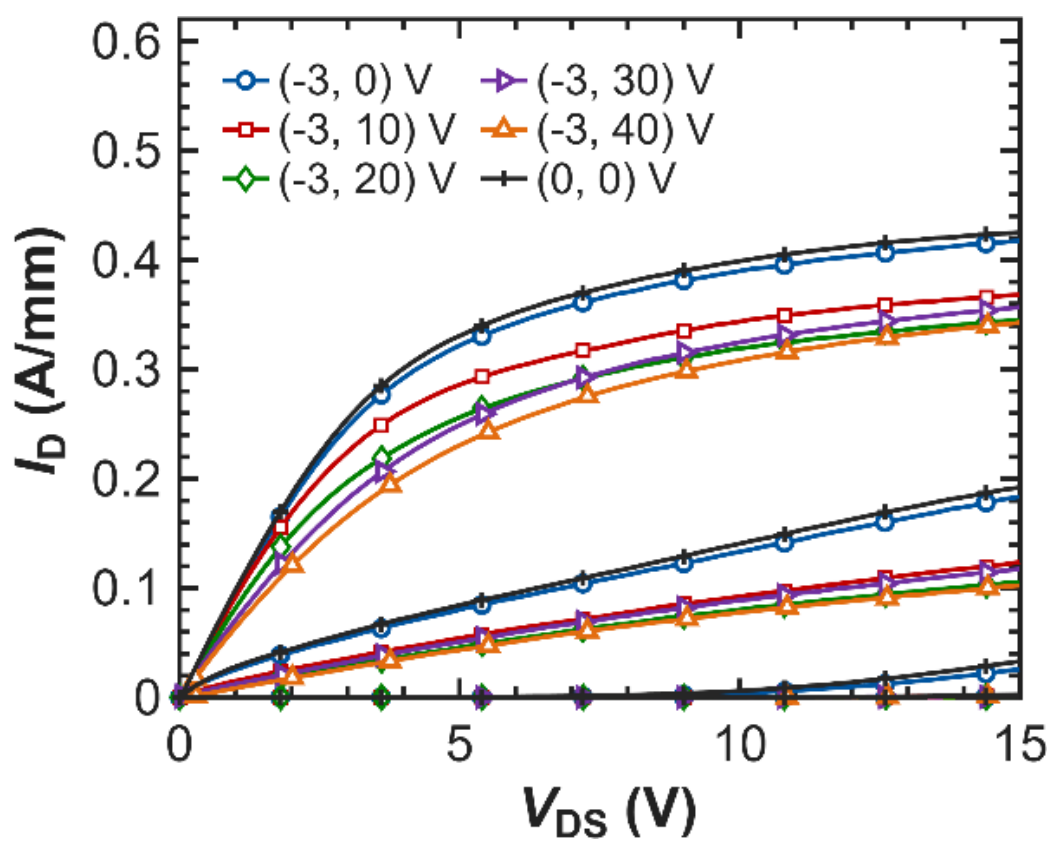

**Figure 5.** Pulsed I-V characteristics of device A at zero quiescent bias and different pinch-off quiescent bias points ($V_{gsq}$, $V_{dsq}$).

Fig. 5 shows the results from pulsed I-V measurement performed with a 50 µs pulse width and 0.1% duty cycle. A peak $I_D$ of 0.425 A/mm was measured at $V_{GS}$ = 2 V and $V_{DS}$ = 15 V for a zero quiescent bias condition. For all the bias conditions used for the pulsed IV measurement, the current collapse in the device remained within an acceptable range. When gate-stress ($V_{gsq}$, $V_{dsq}$) = (-3 V, 0 V) was applied, current reduced by only 1.8% compared to the zero quiescent bias condition ($V_{gsq}$, $V_{dsq}$) = (0 V, 0 V), which shows negligible gate lag. When drain-stress was applied ($V_{gsq}$, $V_{dsq}$) = (-3 V, 10 V), current reduced by 13.4% compared to the zero quiescent bias condition, which indicates electron trapping. However, when $V_{dsq}$ = 20 V was applied, current only dropped by another 5.4%, and after that, for $V_{dsq}$ = 30 V and $V_{dsq}$ = 40 V, current almost remained at the same level with negligible increase in knee walkout. The 40 V drain bias was limited by the tool limit.

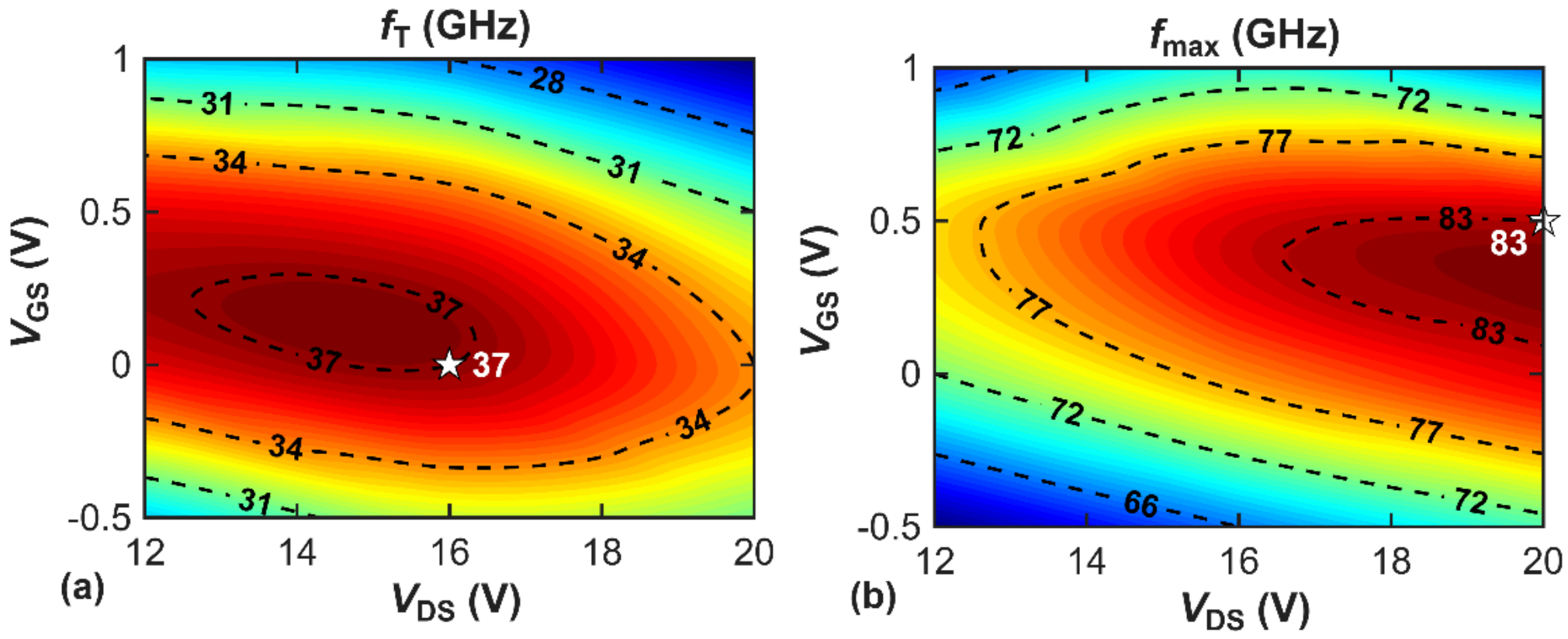

**Figure 6.** Contour plots showing variation in frequency performance of device A with different ($V_{gsq}$, $V_{dsq}$) (a) $f_T$ (b) $f_{max}$

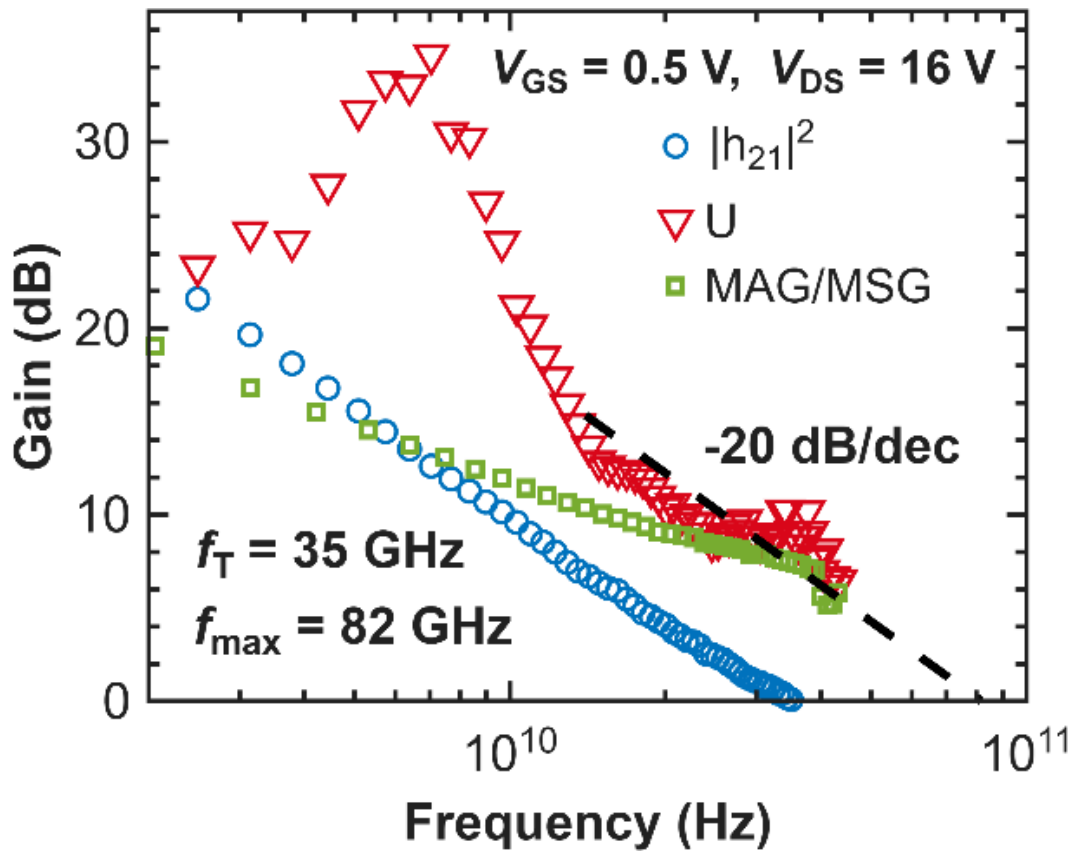

**Figure 7.** $|h_{21}|$, U, and MAG/MSG extracted from S-parameter measurements on device A at ($V_{gsq}$, $V_{dsq}$) = (0.5, 16) V.

S-parameters were measured at room temperature with SOLT calibration followed by on-wafer open and short de-embedding. The device was tested at different bias conditions. Contours for $f_T$ and $f_{max}$ are shown in Fig. 6(a)-(b). The highest measured $f_T$ = 37 GHz was found at ($V_{gsq}$, $V_{dsq}$) = (0 V, 16 V), as shown in Fig. 6(a). The highest measured $f_{max}$ = 83 GHz was found at ($V_{gsq}$, $V_{dsq}$) = (0.5 V, 20 V), as shown in Fig. 6(b). It can be noticed that the peak $f_{max}$ = 83 GHz was obtained at $V_{dsq}$ = 20V. Higher $f_{max}$ can be achieved with increased $V_{dsq}$. The highest $V_{dsq}$ used here was limited by the tool's capability. The overall best combination of $f_T$ and $f_{max}$ was found to be 35 GHz and 82 GHz, respectively, obtained at ($V_{gsq}$, $V_{dsq}$) = (0.5 V, 16 V) as shown in Fig. 7.

**Table I**
**Small-signal parameters**

| **Parameters** | **Value** |
|---|---|
| $R_s$ | 3.61 Ω·mm |
| $R_d$ | 4.43 Ω·mm |
| $R_g$ | 1.5 Ω |
| $C_{gs}$ | 847.6 fF/mm |
| $C_{gd}$ | 112.7 fF/mm |
| $g_{m,ext}$ | 127.5 mS/mm |
| $g_{m,int}$ | 236.2 mS/mm |
| $R_i$ | 1.52 Ω·mm |
| $R_{ds}$ | 271.4 Ω·mm |
| $f_T$ | 35 GHz |
| $f_{T,\text{model}}$ | 35 GHz |
| $f_{max}$ | 82 GHz |
| $f_{max,\text{model}}$ | 82 GHz |

Small-signal parameters were extracted for ($V_{gsq}$, $V_{dsq}$) = (0.5 V, 16 V). The results are shown in Table I. The extrinsic transconductance ($g_{m,ext}$) from small-signal performance is found to be similar to that from DC transfer characteristics. This shows excellent gate control even at very high frequency. The calculated intrinsic transconductance ($g_{m,int}$) in this work is ~236 mS/mm, which indicates that with improvement in ohmic contact resistance, even higher frequency performance can be expected. The capacitance values extracted here are higher than previously reported values for AlGaN channel FETs [17], [19], which is expected due to the reduced distance between the gate and the channel in a recessed gate structure.

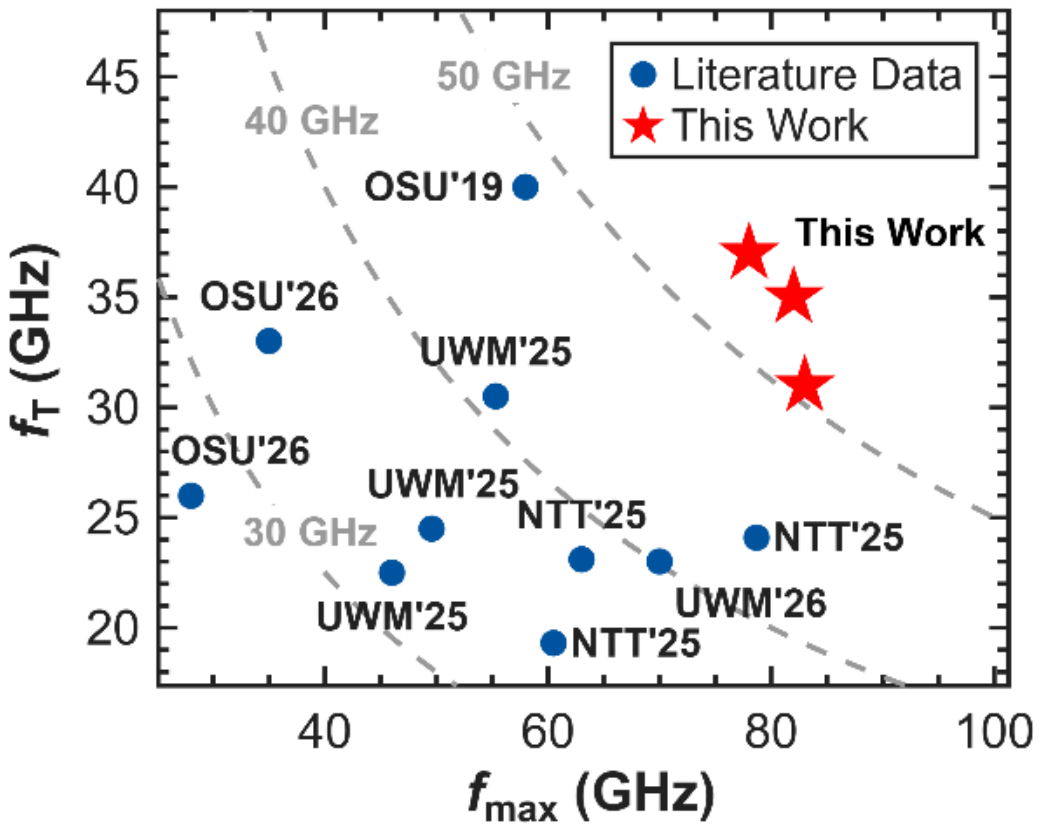

**Figure 8.** Benchmarking plot of $f_T$ and $f_{max}$ for UWBG AlGaN channel HFETs. The dashed lines indicate constant $(f_T.f_{max})^{1/2}$ values.

Fig. 8 benchmarks $f_T$ and $f_{max}$ obtained simultaneously at a single bias point on a device, for AlGaN channel FETs with $f_{max}$ > 25 GHz and for Aluminum composition > 50% [15], [17], [18], [19], [23], [24], [25]. To the authors' best knowledge, the $f_T$ and $f_{max}$ values achieved from the proposed recessed-gate design is the highest combination achieved so far for UWBG AlGaN channel devices.

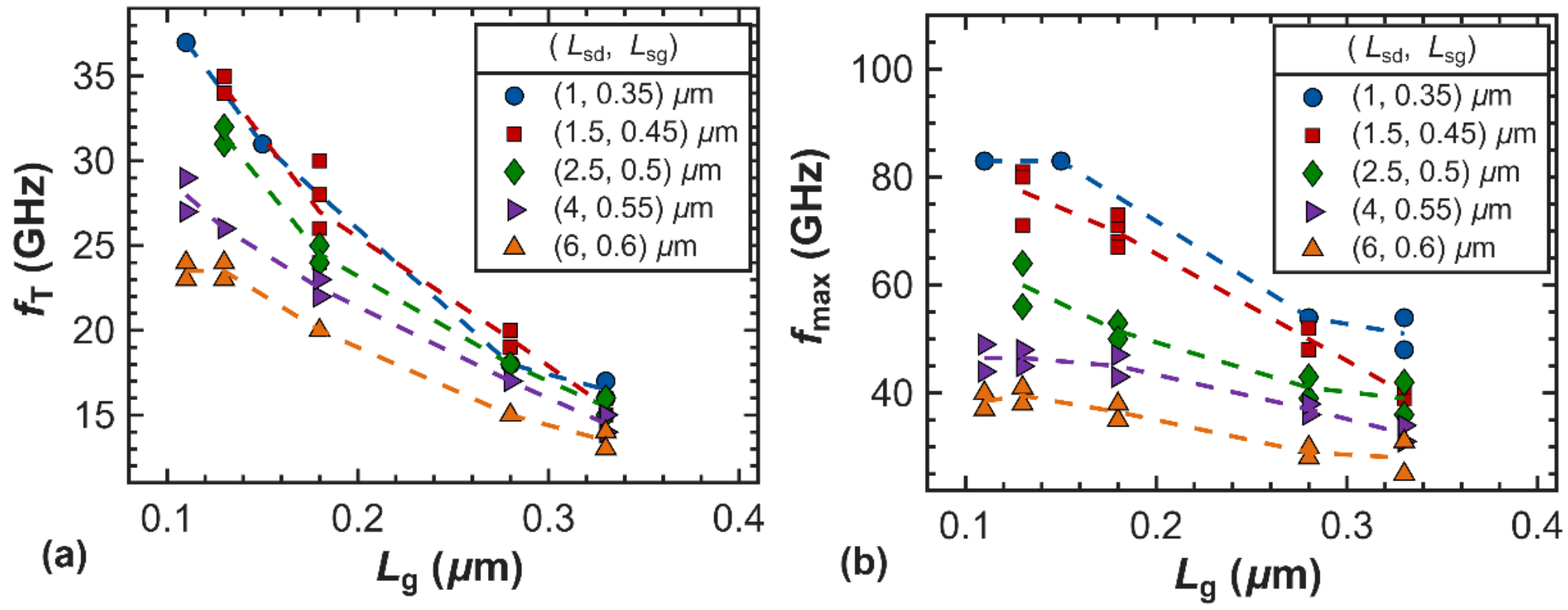

**Figure 9.** Peak frequency performance obtained for different device dimensions (a) $f_T$ (b) $f_{max}$

To further investigate the frequency performance of the recessed gate structure, frequency performance was evaluated for devices with various geometries. Fig. 9(a) shows the variation in $f_T$ with $L_g$ for different $L_{sd}$ and $L_{sg}$. The $f_T$ reduced with an increase in $L_g$, which is due to increased carrier transit time under the gate. Some reductions in $f_T$ can be seen for devices with larger access region, which can be attributed to increased access region resistance. It is noticeable that, for $f_T$ the gate length scaling plays a more prominent role compared to access region scaling. Fig. 9(b) shows the variation in $f_{max}$ with $L_g$ for different $L_{sd}$ and $L_{sg}$. The $f_{max}$ also reduces with increased $L_g$. However, the reduction in $f_{max}$ due to increased access region is noticeably more prominent, as access region resistance plays a significant role in determining the $f_{max}$ of a device.

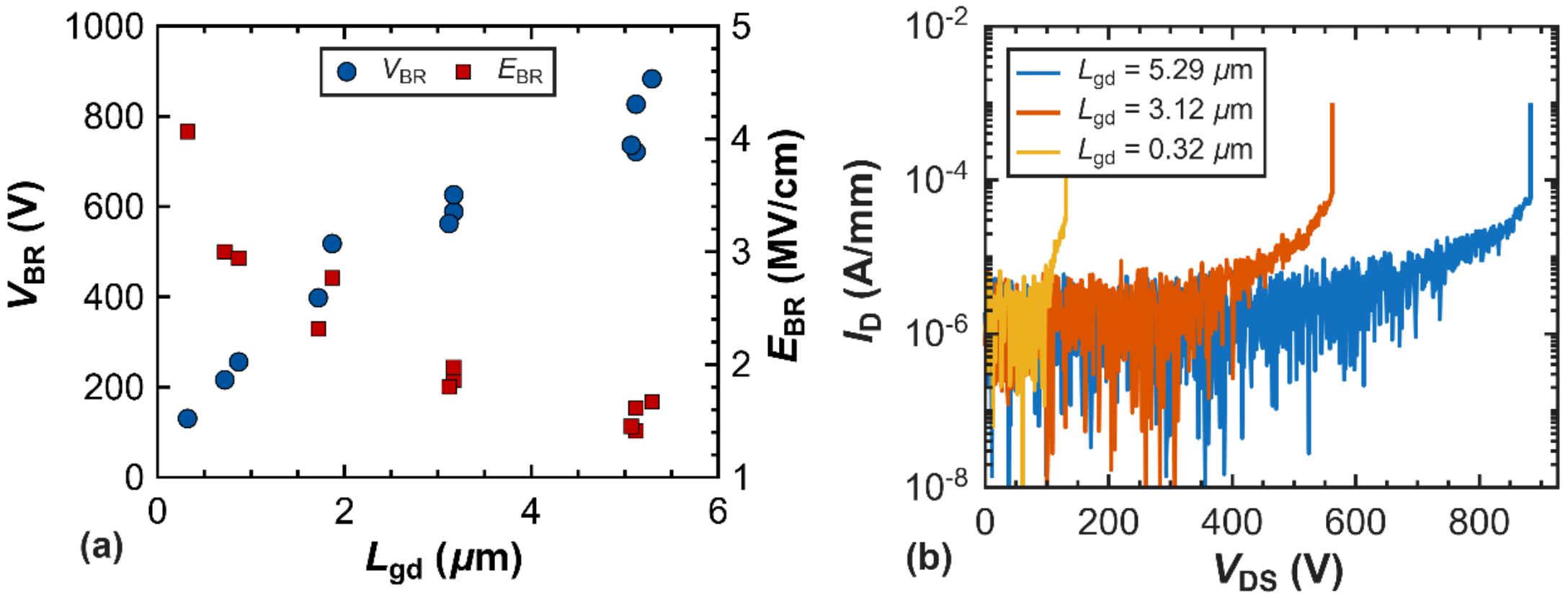

**Figure 10.** (a) Measured three-terminal off-state breakdown voltage and average breakdown field obtained for devices with different $L_{gd}$. (b) Off-state breakdown characteristics for three devices with different $L_{gd}$. The blue curve represents device B.

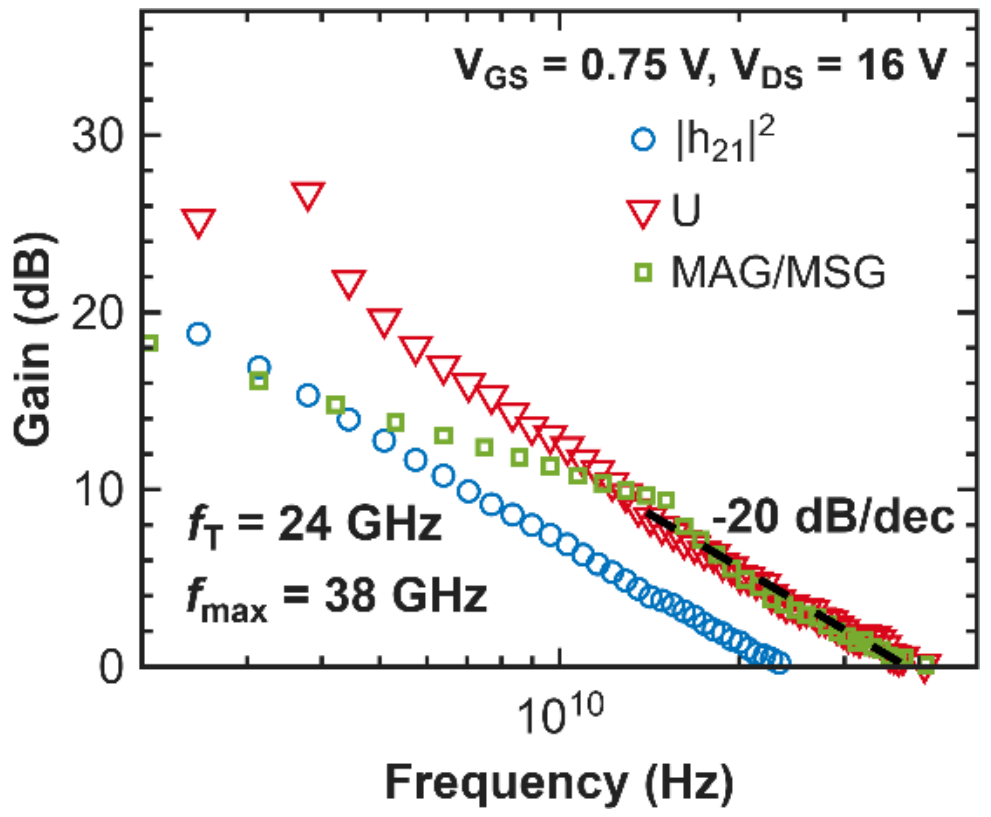

**Figure 11.** $|h_{21}|$, U, and MAG/MSG extracted from S-parameter measurement on device B at ($V_{gsq}$, $V_{dsq}$) = (0.75, 16) V

To evaluate the power handling capability of the devices with recessed-gate structure, three-terminal off-state breakdown was measured with varying $L_{gd}$, as shown in Fig. 10(a)-(b). For the proposed recessed gate structure, the $V_{BR}$ varied from 130 V to 883 V for $L_{gd}$ from 0.32 µm to 5.29 µm, which corresponds to an average breakdown electric field of 4.06 MV/cm for the small $L_{gd}$ device, and 1.67 MV/cm for the large $L_{gd}$ device.

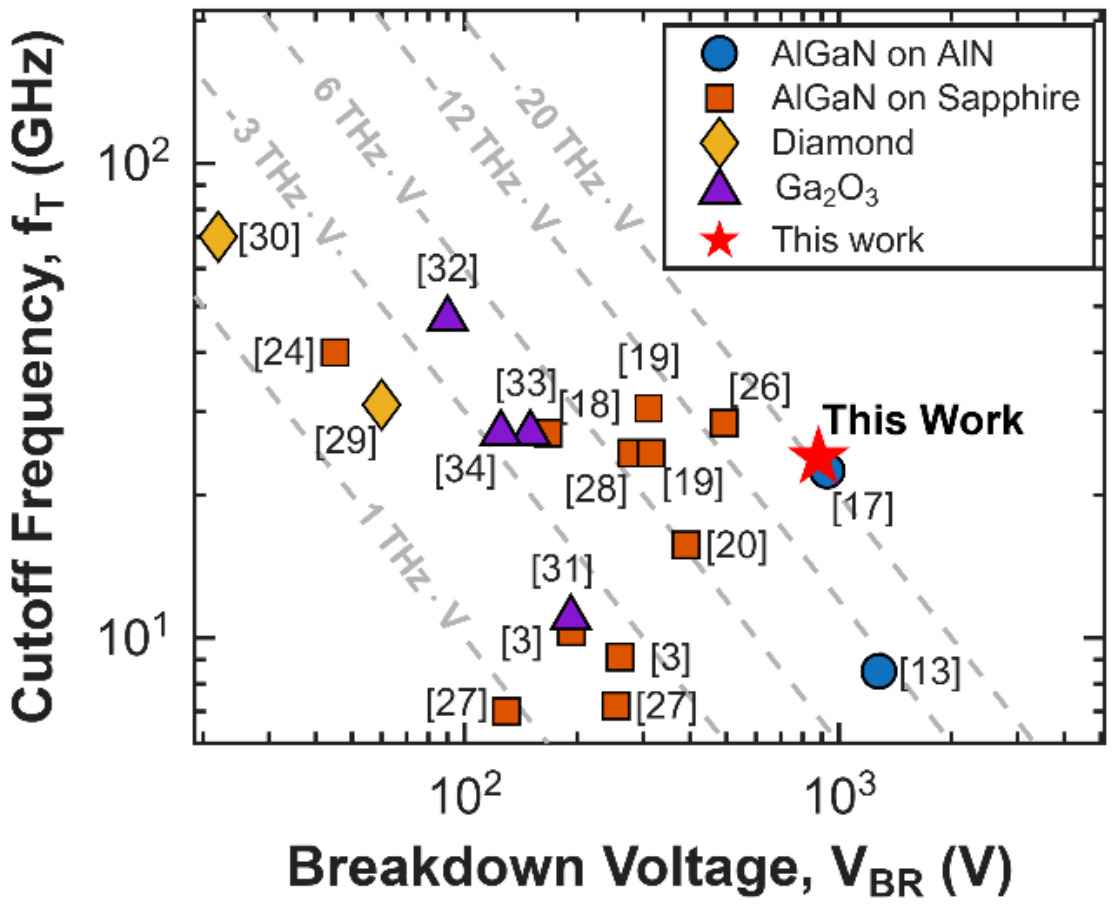

**Figure 12.** Benchmarking plot of $f_T$ and $V_{BR}$ for UWBG channel HFETs.

Device B, which exhibited a $V_{BR}$ of 883 V, demonstrated a peak $f_T$ of 24 GHz, corresponding to a high Johnson's figure of merit (JFOM) of 21.19 THz·V. Fig. 11 shows the data from small-signal measurement for this device. A benchmark comparing $f_T$ and $V_{BR}$ across various UWBG materials is shown in Fig. 12 [3], [13], [18], [17], [19], [20], [24], [26], [27], [28], [29], [30], [31], [32], [33], [34]. As observed in Fig. 12, previous reports of AlGaN channel RF HEMTs on sapphire substrates were limited to breakdown voltages of < 500 V when $f_T$ was greater than 5 GHz. Although higher voltages > 1000 V were reported for RF HEMTs on sapphire [27], frequency performance remained low for those devices, with $f_T$ < 5 GHz due to increased resistances from larger access region dimensions. Higher breakdown voltages with high frequency performance and, consequently, high JFOM were achieved for devices fabricated on AlN substrate. With the proposed recessed gate device design, we were able to obtain both high breakdown voltage and high $f_T$ on sapphire substrate, which provided a similar JFOM compared to the previously recorded highest JFOM for UWBG materials. This high JFOM on sapphire substrate was achievable due to improved frequency performance with the proposed recessed gate structure.

In this work, we have demonstrated a recessed-gate passivated $Al_{0.60}Ga_{0.40}N$ channel HEMT which showed an $f_{max}$ of 82 GHz and an $f_T$ of 35 GHz with reasonable dispersion control. This excellent frequency performance was achieved due to the simultaneous presence of low on-resistance and high transconductance in the device resulting from the recessed-gate structure. Additionally, a high JFOM of 21.19 THz·V and a high electric field of >3 MV/cm have been obtained from these devices. The devices also exhibited a high $I_{ON}/I_{OFF}$ ratio by maintaining low drain and gate leakage currents. These findings underscore that AlGaN channel FETs can be a great candidate for high-power RF applications from X- to Ka-band.

**Acknowledgments**

This work was supported in part by NSF ASCENT (award No. ECCS 2328137), ARPA-E Award DE- AR0001825, NSF Career Award ECCS 2441140, AFOSR Grant FA9550-23-10501, and FA9550-24-1-0159.